\documentclass{IEEEtran}

\usepackage{graphicx} % Required for inserting images
\usepackage{xcolor, bm, amsmath}
\usepackage{amsfonts, amssymb}
\usepackage{mathtools}
\usepackage{enumitem}
\usepackage{comment}
\usepackage{lipsum}
\usepackage{soul}
\usepackage{balance}
\usepackage{booktabs}
\usepackage{tabularx}

\usepackage{array}
\newcolumntype{L}[1]{>{\raggedright\let\newline\\\arraybackslash\hspace{0pt}}m{#1}}
\newcolumntype{C}[1]{>{\centering\let\newline\\\arraybackslash\hspace{0pt}}m{#1}}
\newcolumntype{R}[1]{>{\raggedleft\let\newline\\\arraybackslash\hspace{0pt}}m{#1}}

\usepackage[acronym,style=alttree,nomain,shortcuts,nonumberlist,nogroupskip,nopostdot]{glossaries}
 
\newacronym{irs}{IRS}{intelligent reflecting surface}
\newacronym{ula}{ULA}{uniform linear array}
\newacronym{cr}{CR}{challenge response}
\newacronym{csi}{CSI}{channel state information}
\newacronym{cdf}{CDF}{cumulative distribution function}
\newacronym{pla}{PLA}{physical layer authentication}
\newacronym{awgn}{AWGN}{additive white Gaussian noise}
\newacronym{dft}{DFT}{discrete Fourier transform}
\newacronym{glrt}{GLRT}{generalized likelihood ratio test}
\newacronym{lrt}{LRT}{likelihood ratio test}
\newacronym{ml}{ML}{maximum likelihood}
\newacronym{mimo}{MIMO}{multiple-input multiple-output}
\newacronym{bs}{BS}{base station}
\newacronym{fa}{FA}{false alarm}
\newacronym{md}{MD}{missed detection}
\newacronym{ofdm}{OFDM}{orthogonal frequency division multiplexing}
\newacronym{ue}{UE}{user equipment}
\newacronym{snr}{SNR}{signal-to-noise ratio}
\newacronym{pdf}{pdf}{probability density function}
\newacronym{map}{MAP}{maximum a posteriori probability}
\newacronym{kl}{KL}{Kullback-Leibler}
\newacronym{det}{DET}{detection error tradeoff}
\newacronym{gnb}{gNB}{Next Generation Node B}
\newacronym{iot}{IoT}{Internet of things}
\newacronym{pcc}{PCC}{partially controllable channel}

\title{Hybrid Channel- and Coding-Based Challenge-Response Physical-Layer Authentication}

 \author{ \IEEEauthorblockN{Laura~Crosara,~\IEEEmembership{Member,~IEEE}, Mahtab Mirmohseni,~\IEEEmembership{Senior Member,~IEEE},  \\and Stefano~Tomasin,~\IEEEmembership{Senior Member,~IEEE}
\thanks{Manuscript received --; revised -- and -- accepted --. Date of publication --; date of current version --.} 
\thanks{Corresponding author: L. Crosara. }
\thanks{L. Crosara and S. Tomasin are with the Department of Information Engineering, Universit\`a degli Studi di Padova, Padua 35131, Italy. (email:  {\{laura.crosara.1@phd., stefano.tomasin@ \}unipd.it }).}
\thanks{Mahtab Mirmohseni is with the Institute for Communication Systems, University of Surrey, Guildford, Surrey GU2 7XH, United Kingdom (email: m.mirmohseni@surrey.ac.uk).}
\thanks{This work is supported by the project ISP5G+ ( CUP D33C22001300002), which is part of the SERICS program (PE00000014) under the NRRP MUR program funded by the EU-NGEU" and by the European Commission through the Horizon Europe/JU SNS project ROBUST-6G (Grant Agreement no. 101139068). %\hl{ack to COST?} 
}
}}

\begin{document}
\sloppy
\maketitle

\begin{abstract}
%This letter investigates the use of a \ac{pcc} to support \ac{pla} mechanisms, with coding and channel-based \ac{pla}. The primary objective is to authenticate the communication between a legitimate transmitter and receiver (the verifier) by exploiting the configurable properties of the channel.%, which are under the receiver's control as a channel metric. The verifier changes the physical properties of the electromagnetic environment and expects to receive a properly modified signal from the device under verification. We derive the number of secret bits including contributions from coding and additional randomness introduced by the \ac{pcc}. Our results show that the combined use of both \ac{pla} mechanisms significantly decreases the attacker's success probability.
This letter proposes a new physical layer authentication mechanism operating at the physical layer of a communication system where the receiver has partial control of the channel conditions (e.g., using an intelligent reflecting surface). We aim to exploit both instantaneous channel state information (CSI) and a secret shared key for authentication. This is achieved by both transmitting an identifying key by wiretap coding (to conceal the key from the attacker) and checking that the instantaneous CSI corresponds to the channel configuration randomly selected by the receiver. We investigate the trade-off between the pilot signals used for CSI estimation and the coding rate (or key length) to improve the overall security of the authentication procedure.
%
%This letter investigates the authentication problem at the physical layer, a security mechanisms aiming at authenticating the communication between a legitimate transmitter and receiver by exploiting the configurable properties of the channel, which are under the receiver's control. To assess the performance of this hybrid authentication scheme we measure the quantity of information (in bits) that remains secret to the attacker and protects the authentication procedure.
\end{abstract}

\glsresetall

%Limit: 5500 words, 7 pages

\IEEEpeerreviewmaketitle

\section{Introduction} 
Source authentication mechanisms aim to determine whether a message received truly comes from the declared sender or has been falsified by an attacker. For \ac{pla}, two main approaches are available, denoted here as coding-based \ac{pla} and tag-based \ac{pla}.

The {\em coding-based \ac{pla}} mechanism involves the use of wiretap coding \cite{lai2009authentication}. The verifier and the legitimate transmitter share a key to remain secret to the attacker. The key is used as the secret message in a wiretap-coding transmission from the legitimate transmitter. The verifier checks that the received secret message corresponds to the shared key.

The {\em tag-based} \ac{pla} mechanism~\cite{10.1007/3-540-39568-7_32} is based on the \ac{csi} and includes the two phases of acquisition and verification. In the acquisition phase, the verifier estimates the \ac{csi} from the legitimate source. In the verification phase, upon reception of a message, the verifier estimates the \ac{csi} and compares it to that obtained in the acquisition phase. The message is considered authentic when the two estimates match.
Tag-based \ac{pla} has been applied to technologies such as \ac{ofdm}, \ac{mimo} \cite{6204019, 6804410}, and underwater acoustic networks \cite{Tomasin-twc18-1}. The Neyman-Pearson test \cite{MaurerAuthTh} and machine learning classifiers \cite{8533399} have been used for verification. See \cite{Tomasin-proc} and \cite{9279294} for a comprehensive review.

Recent evolution of \ac{pla} leverages {\em \acp{pcc}}, i.e.,  channels partially controlled by the receiver through a modification (configuration) of the signal propagation environment, e.g., using a \ac{irs} to steer wireless signals. Even this mechanism includes two phases. In the first, the verifier obtains estimates of the channel associated with all possible configurations of the propagation environment. In the second, the verifier sets a random configuration and, upon receiving a message, estimates the channel and checks its consistency with the channel predicted from the estimates of the first phase~\cite{tomasin2022challenge}. The mechanism is denoted as {\em channel-based \ac{cr}-\ac{pla}} as it mimics the \ac{cr} authentication mechanisms using cryptographic primitives.

In this letter, we propose a hybrid PLA mechanism that combines channels- and coding-based \ac{pla} mechanisms. The two mechanisms intersect in their use of pilot and data symbols. An increased number of pilot symbols benefits the channel-based mechanism, while a higher data-to-pilot ratio favors the coding-based mechanism. Thus, we analyze the trade-off between the amount of transmitted data and pilot symbols. Subsequently, we assess the security of the channel-based, coding-based, and hybrid \ac{pla} mechanisms in terms of bits that the attacker must know to deliver a successful attack. Another key contribution of this letter is the introduction of the following features for existing schemes: (a) the consideration of multiple channel configurations in \ac{cr}-\ac{pla}, and (b) the incorporation of finite-length coding in coding-based \ac{pla}. Additionally, for the hybrid \ac{pla} analysis, we extend the existing coding-based \ac{pla} framework to account for block-fading channels. Our study demonstrates that the hybrid \ac{pla} mechanism offers improved authentication capabilities over each existing mechanism, particularly when the attacker's channel \ac{snr} is significantly lower than that of the nominal channel.

The rest of this letter is organized as follows. Section~\ref{sec:sysmod} presents the system model. The proposed mechanism is introduced in Section~\ref{hcrpla}. Section~\ref{sec:analysis} provides the analytical framework for evaluating the security properties of the system. Then, Section~\ref{sec:numres} presents the numerical results. Finally, Section~\ref{sec:concl} concludes the letter.

\section{System Model}\label{sec:sysmod}

\begin{figure}
    \centering
    \includegraphics[width=0.95\columnwidth]{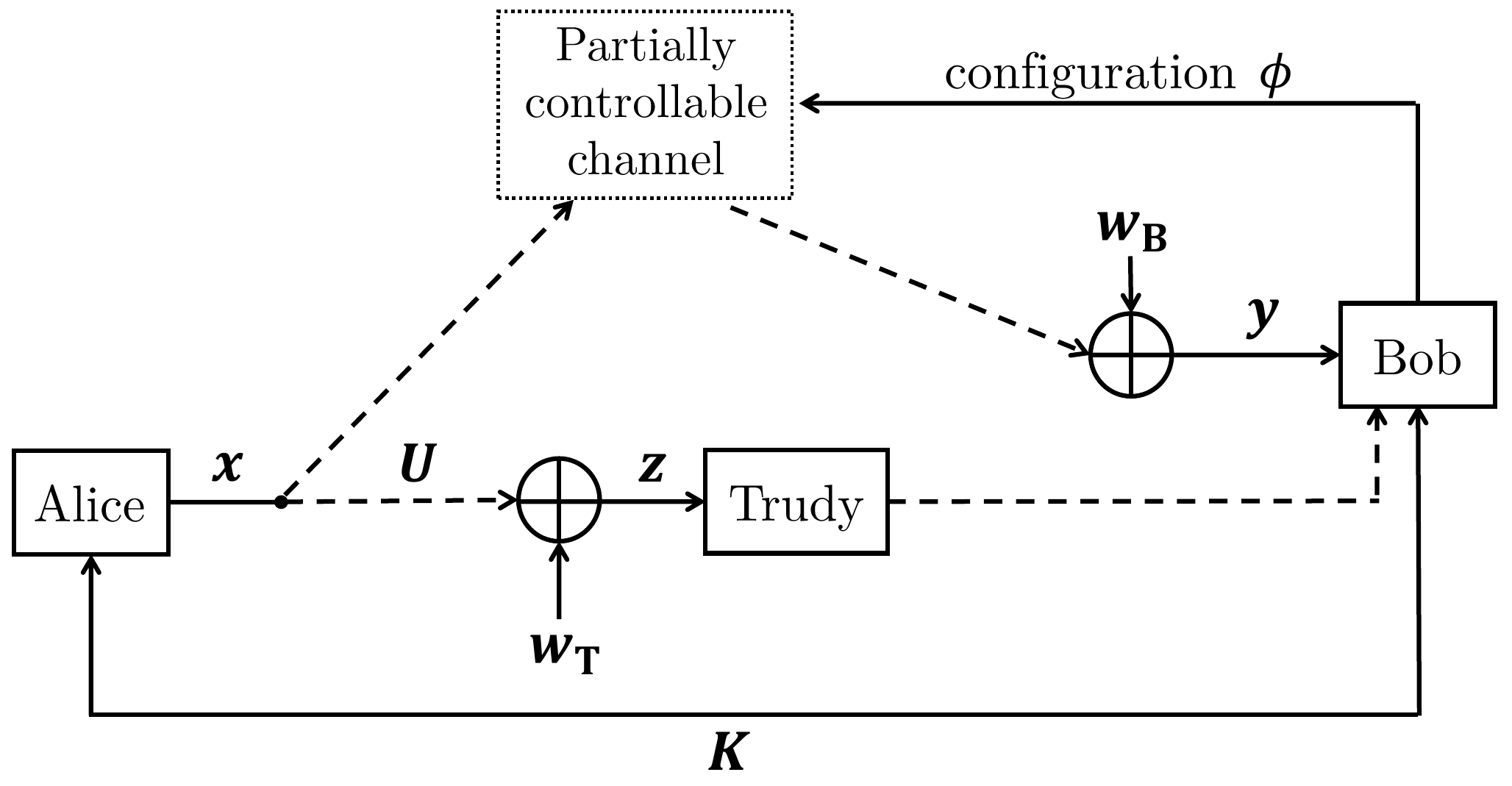}
    \caption{Considered communication model. }%\hl{Quando Trudy trasmette non c'e' rumore a Bob, il disegno va aggiornato  -- c'e' una virgola dopo configuration nella figura?}}
    \label{fig:ul_2}
\end{figure}

We consider the uplink wireless transmission scenario of Fig.~\ref{fig:ul_2}, where a receiver Bob aims at authenticating messages coming from a legitimate transmitter Alice. An intruder Trudy, in turn, aims at impersonating Alice, transmitting fake (spoofing) messages to Bob.
We consider a single-antenna setup for Alice, Bob, and Trudy, while an extension to a multi-antenna scenario is possible. We assume that the signal transmitted by Alice $x^{(t)}$ at discrete time $t$ is zero-mean Gaussian with unit power.

\subsection{Channel Model}

A \textit{\ac{pcc}} supports the communication between Alice and Bob \cite{tomasin2022challenge}. Bob can adjust certain channel properties by selecting a channel \textit{configuration} $\phi$ via a dedicated channel that Trudy cannot access. For example, a \ac{pcc} can be implemented with an \ac{irs}, which reflects radio signals with adjustable phases. A specific configuration is achieved by setting the phase shift values of the \ac{irs} elements. However, in this letter, we consider a generic controllable channel, without explicit reference to IRSs.

At symbol time $t$ of the considered single-input-single-output (SISO) \ac{awgn} model, let $h(\phi)$ be the (real) amplitude of the channel from Alice to Bob when configuration $\phi$ is used and $w^{(t)}_{\mathrm{B}}$ be the complex circularly symmetric \ac{awgn} term with zero mean and variance $\sigma^2_{\rm B}$. For the transmission of $n$ complex symbols $x^{(t)}$, $t = 1, \ldots, n$, the input-output relation with \ac{irs} configuration $\phi$ is\footnote{We assume here the phase of the resulting channel cannot be controlled by Bob as it is also related to synchronization issues that are not easily controllable.} %of frame $k$ from Alice to Bob is 
\begin{equation}\label{eq1}
y^{(t)} = h(\phi)e^{j\theta} x^{(t)}+w^{(t)}_{\mathrm{B}}, \quad t=1, \ldots, n,
\end{equation}
where $\theta$ is a random phase introduced by the channel, independent from $\phi$. The legitimate channel \ac{snr} is maximized when the communication-optimal configuration is selected. When Bob chooses a configuration that deviates from the communication-optimal one, the legitimate channel \ac{snr} decreases. We assume that the selected configuration is close to the communication-optimal one, ensuring that the channel’s amplitude $h(\phi)$ satisfies
\begin{equation}\label{eq:chconst}
  h_{\rm MIN} \leq h(\phi) \leq h_{\rm MAX}\,,
\end{equation}
where $h_{\rm MAX}$ is the channel obtained with the communication-optimal (maximum \ac{snr}) configuration, while $ h_{\rm MIN}$ is the minimum amplitude achieved with the configuration leading the worse channel.
We further assume that for each $h\in [ h_{\rm MIN},  h_{\rm MAX}]$ there is at least one configuration $\phi$ that satisfies $h(\phi) = h$.

The input-output relation of the Alice-Trudy channel, which is not controllable by Bob, is 
\begin{equation}
z^{(t)} = Ue^{j\theta'} x^{(t)}+w^{(t)}_{\mathrm{T}}, \quad t=1, \ldots, n, 
\end{equation}
where $w^{(t)}_{\mathrm{T}}$ follows a zero-mean Gaussian distribution with variance $\sigma^2_{\rm T}$, and $\theta'$ is a random phase introduced by the channel. We define %the \ac{snr} on the legitimate and attacker channel as 
$\Lambda_{\rm B} = 1/\sigma^2_{\rm B}$ and $\Lambda_{\rm T} = U^2/\sigma^2_{\rm T}$.%, respectively.

\subsection{Attack Model}
Trudy does not know $\phi$ and, consequently, $h(\phi)$, but is aware of $h_{\rm MAX}$ and $h_{\rm MIN}$. To consider the worst-case scenario we assume that Trudy can induce any channel estimate $a$ at Bob and, when Trudy attacks, the received signal is noiseless, thus under attack
\begin{equation}
y^{(t)} = a x_{\rm T}^{(t)}, \quad t=1, \ldots, n.
\end{equation}
where $x_{\rm T}^{(t)}$ is the symbol transmitted by Trudy at time $t$.

%Trudy generates $a$ uniformly at random in $[h_{\rm MIN}, h_{\rm MAX}]$ and independently on each frame.

\section{Hybrid CR-PLA Mechanism}\label{hcrpla}
The proposed {\em hybrid \ac{cr}-\ac{pla}} mechanism combines \textit{channel-based \ac{cr}-\ac{pla}} mechanisms with {\em coding-based \ac{pla}} techniques. Here we introduce two new elements in the two approaches. For \ac{cr}-\ac{pla} we consider that multiple configurations are used to transmit a single message and for coding-\ac{pla} we consider codes of finite length.

\subsection{Channel-Based CR-PLA}\label{sec:CHCRPLA}
For the \textit{channel-based \ac{cr}-\ac{pla}} (CH-CRPLA), Bob exploits only the amplitude of the channel $h(\phi)$ which in the following is also denoted as \ac{csi}. The CH-CRPLA mechanism \cite{tomasin2022challenge} includes two phases. 

In the {\em association} phase Alice transmits several pilot signals over the \ac{pcc}, each corresponding to a different configuration chosen by Bob. For each configuration $\phi$, Bob estimates the \ac{csi} $h(\phi)$, and stores both the estimated \acp{csi} and the corresponding configurations. The configurations used in this phase are such that Bob obtains estimates of the \ac{csi} also for any other configuration. It is assumed that in this phase Trudy is not transmitting or that the pilot signals are authenticated with other mechanisms; we also assume that the \ac{csi} estimated by Bob is perfect.

In the {\em verification} phase, Alice first splits the message into $F$ frames, indexed by $k=1,\hdots,F$, each comprising $n$ symbols, indexed by $t=1,\hdots,n$. In detail, $\alpha n$ are pilot symbols, and the remaining $n'= (1-\alpha)n$ are used to transmit $b_{\rm M}$ bits of information (the message). Then, Bob transmits each frame with a different \ac{pcc} configuration $\phi_k$, chosen such that \ac{csi} $h(\phi_k)$ is uniformly distributed in $[h_{\rm MIN}, h_{\rm MAX}]$ and independent for each frame $k$. Bob leverages the $\alpha n$ pilots to estimate the \ac{csi} for each frame $k$ as 
\begin{equation}
\hat{h}(\phi_k) = \frac{1}{\alpha n} \sum_{t=1}^{\alpha n} \frac{y_k^{(t)}}{x_k^{(t)}} \sim {\mathcal N}(h(\phi_k), \sigma^2_h),
\end{equation}
with $\sigma^2_h = \sigma_{\rm B}^2/(\alpha n)$. Then, Bob checks the consistency of the estimated \ac{csi} with the expected \ac{csi} $h(\phi_k)$. To this end, we use the Neyman-Pearson optimal test function 
\begin{equation}\label{testeq}
L = \frac{1}{\sqrt{2F}} \left[\sum_{k=1}^F \left(\frac{[\hat{h}(\phi_k)- h(\phi_k)]^2}{\sigma_{h}^2}\right)  - F\right],
\end{equation}
and the message is considered authentic if $L>\tau$. We note that due to a specific realization of the noise, we may have $L<\tau$, and in this case, we have a \ac{fa} event. The threshold $\tau$ is chosen to achieve the desired \ac{fa} probability. 

Correspondingly, Trudy may change the attack channel at each frame, thus using channels $\bm{a} = [a_1, \ldots, a_F]$.

We remark that the CH-\ac{cr}\ac{pla} protocol does not rely on any pre-shared secret.

%% qui c'era hMIN hMAX

\subsection{Coding-Based PLA}\label{sec:CDPLA}

In the \textit{coding-based \ac{pla}} (CD-PLA) mechanism~\cite{lai2009authentication} Alice and Bob share a key $K$ of $b_{\rm key}$ bits, potentially derived from the \ac{csi} through a secret key agreement procedure. Then, a wiretap code is used to encode the key as the confidential information to be concealed from Trudy, while the message of $b_{\rm M}$ bits provides the random part. Leveraging the results on secrecy capacity \cite[Ch. 3]{bloch2011physical}, we assume here that the error-correcting code has Gaussian codewords of $nF$ symbols that provides $x^{(t)}$. To confirm the authenticity of the message, Bob decodes the received signal and verifies that the decoded key is $K$.

%When Alice is transmitting with long enough codewords ($nF \rightarrow \infty$) and, 
When only CD-PLA is used, the \ac{pcc} configuration is kept fixed at $h_{\rm MAX}$ to maximize the data rate on the Alice-Bob channel.
The mutual information $I(\cdot;\cdot)$ between $x$ and $y$ is
\begin{equation}
    I(x;y) = \log_2(1 + h_{\rm MAX}^2\Lambda_{\rm B})\,.
\end{equation}
As $h_{\rm MAX}$ is known to Bob from the association phase, no pilot signals are transmitted with the message in the verification phase. 

Considering the rate reduction due to the use of finite-length codes in the presence of a Gaussian channel \cite[eqs.(1), (293)]{5452208}, the information rate between Alice and Bob is
\begin{equation}
    R = I(x;y) - \sqrt{\frac{\Lambda_B (\Lambda_B+2)\log^2_2 e}{(\Lambda_B +1)^2 nF}} Q^{-1}(p_{\rm FA})\,,
\end{equation}
where $Q^{-1}(\cdot)$ is the inverse of the Q-function $Q(\cdot)$.

Alice can transmit a message of size $b_{\rm M}$ bits with decoding error probability $p_{\rm FA}$ at Bob when
\begin{equation}\label{eq:lai1}
nF R > b_{\rm M} + b_{\rm key}.
\end{equation}
To ensure also that Trudy does not obtain any information on the key we must have \cite{lai2009authentication} 
\begin{equation}\label{eq:lai2}
nF [R -I(x;z)] > b_{\rm key}.
\end{equation}
where we upperbounded the Trudy-Bob rate to $I(x;z)= \log_2(1 + \Lambda_{\rm T})$ ignoring the rate loss due to the finite-length coding.

Considering a key of maximum length, we define 
\begin{align}\label{eq:bk12}
    b_{\rm{key}, 1}^{\rm CD} &= nF R - b_{\rm M},\\ 
    b_{\rm{key}, 2}^{\rm CD} &= nF\,[R - I(x;z) ].
\end{align}
Then, the number of secret bits of the key with the CD-PLA method is
\begin{equation}
    b_{\rm key}^{\rm CD} = \max\{0, \min\{b_{\rm{key}, 1}^{\rm CD} , b_{\rm{key}, 2}^{\rm CD} \}\}.
\end{equation}

\subsection{Hybrid CR-PLA}
We now present the novel \textit{hybrid CR-PLA} (H-CRPLA) mechanism, which combines CH-\ac{cr}\ac{pla} and CD-\ac{pla}.

Also in H-CRPLA, as for CH-\ac{cr}\ac{pla}, we have the two phases of \textit{association} and \textit{verification}, the latter split into frames. At each frame, a random i.i.d \ac{pcc} configuration is selected by Bob. However, each frame includes both the wiretap codeword, computed as in CD-\ac{pla} and pilot symbols for the \ac{csi} estimate at Bob. In particular, a codeword of $(1-\alpha)Fn$ Gaussian symbols is obtained from the secret key of $b_{\rm key}$ bits and the message of $b_{\rm M}$ bits. This codeword is split into $F$ parts, each with $n'=(1-\alpha)n$ symbols, and each part is transmitted in a different frame, each including $\alpha n$ pilot symbols.

Upon reception of a message, Bob performs two authentication checks: the \ac{csi} estimated in each frame should match the expected one using \eqref{testeq} (as in channel-based \ac{cr}-\ac{pla}) and the decoded key should be $K$ (as in the coding-based \ac{pla}). A failure in either of the two checks leads to a rejection of the received message as non-authentic. 

An analysis of the security properties of H-CR-PLA is provided in the next Section.

%__________________________________________________________________________________________________________________________________________________________________________
\section{Security Analysis}\label{sec:analysis}

To analyze the security of H-CRPLA, we consider the contributions of CH-CRPLA and CD-PLA. In particular, for both mechanisms, we obtain the number of secret bits the attacker must know to get a successful attack.

\subsection{Security from Channel-Based CR-PLA}
 
For the \ac{cr}-\ac{pla} authentication check within H-CRPLA, we first note that as $F \rightarrow \infty$, the test variable $L$, defined in \eqref{testeq}, tends to a standard normal distribution, and the \ac{fa} probability associated to this test is 
\begin{equation}
p_{\rm FA}^{\rm CH-CRPLA} = \mathbb{P}(L>\tau|\mathcal H_0) \rightarrow Q(\tau).
\end{equation}
Hence (asymptotically) the \ac{fa} probability does not depend on $F$ and the threshold $\tau$ can be set to obtain a given (small) $p_{\rm FA}$. Moreover, the \ac{fa} probability does not depend on the statistics of vector $\hat{\bm{h}} = [\hat{h}(\phi_1), \hdots, \hat{h}(\phi_F)]$.

The CR-PLA test is passed when satisfying \eqref{testeq}, i.e., when $\hat{\bm{h}}$ falls within the hyper-sphere (in a space of size $F$) defined, from \eqref{testeq}, by the following inequality
\begin{equation}\label{hypersph}
  \sum_{k=1}^F [\hat{h}(\phi_k) - h(\phi_k)]^2  \leq (\sqrt{2F}\tau + F)\sigma_{h}^2.
\end{equation}
The hypersphere is centered in $\bm{h}= [h(\phi_1), \hdots, h(\phi_F)]$, with radius
$R_{\rm s} = [(\sqrt{2F}\tau + F)\sigma_{h}^2]^{1/2}$ and volume
\begin{equation}
V_{\rm s} = \frac{\pi^{F/2}}{\Gamma\left(\frac{F}{2} +1\right)} R_s^F,
\end{equation}
where $\Gamma(\cdot)$ is the Euler's Gamma function.

\paragraph*{Defense Strategy}
For choice of the configurations $\phi_k$, $k=1,\hdots,F$, we first observe that the values taken by $\bm{h}$ lay instead in $2^F$ hypercubes (in a space of size $F$) with edges of length $E = h_{\rm MAX}-h_{\rm MIN}$ and volume
\begin{equation}
V_{\rm c} = 2^F E^F.
\end{equation}
Then, Bob chooses $\bm{h}$ uniformly at random in the hypercube; this is roughly equivalent to choosing  $\bm{h}$ such that any point in the hypercube has the same probability of belonging to the hyper-sphere of the decision.

\paragraph*{Attack Strategy} Since all points in the hypercube are equally probable, the best attack for Trudy is to randomly choose a point in the hypercube. Thus, Trudy generates the entries of its attack vector   $\bm{a} = [a_1,\hdots, a_F]$ uniformly at random. 

\paragraph*{Success Probability of Attacks}
Trudy succeeds in her attack when satisfying \eqref{hypersph}. The success probability of the attack is then the ratio of the volume of the hypersphere and the volume of the hypercube upper-bounded by $1$, i.e.,\footnote{For large $F$ the Stirling approximation may be used and compute $\log_2 P_{\rm succ}$ instead of $P_{\rm succ}$ and avoid numerical problems.}
\begin{multline}\label{eq:psucc}
P_{\rm succ} = \min \Bigg\{1\,  ,\;  \frac{V_s}{V_c} = \frac{1}{\Gamma\left(\frac{F}{2} +1\right)}\\ \times \left[\frac{\sqrt{\pi}(\sqrt{2F}\tau + F)^{1/2}\sigma_{h}}{2(h_{\rm MAX}-h_{\rm MIN})}\right]^F\Bigg\},
\end{multline}
where the minimum ensures that $P_{\rm succ} \leq 1$ and the equality is achieved when $V_s\geq V_c$. In \eqref{eq:psucc} we assumed that the radius of the sphere is much smaller than the edge length of the hypercube. This is achieved when $R_s \ll h_{\rm MAX}-h_{\rm MIN}$. This allows us to consider the sphere as entirely contained within the hypercube, neglecting any boundary effects.

Notably, the probability of success $P_{\rm succ}$ is the same as obtaining a specific realization from a random extraction of $b_{\rm ch}$ i.i.d.  bits, with equal probability for each value $\{0,1\}$. Thus, the success probability is the same as finding a random binary key with length
\begin{equation}\label{eq:bch}
b_{\rm ch} = -\log_2 P_{\rm succ}.
\end{equation}
We denote this key as the {\em equivalent CH-\ac{cr}\ac{pla} key}. 

\subsection{Security of Coding-Based PLA}

We now extend the result of \cite{lai2009authentication} to the case of block fading channels and finite-length coding. 
For the considered \ac{awgn} channel, at frame $k$ we have 
\begin{equation}\label{eq:Ixy}
    I_k(x;y) =  \log_2\left( 1+h(\phi_k)^2\Lambda_{\rm B} \right).
\end{equation}
Since $x$ is zero-mean Gaussian, $h(\phi_k)=h_{\rm MAX}$ maximizes the mutual information at frame $k$, $I_k(x;y)$, while the randomness on $\phi_k$ reduces $I_k(x;y)$. 

Leveraging on the results of \cite{6404740}, the average information rate between Alice and Bob is
\begin{equation}
    \bar{R} = \mathbb{E}[I_k(x;y)] - \sqrt{\frac{V}{n'F}} Q^{-1}(p_{\rm FA}^{\rm CD-PLA})\,,
\end{equation}
where the average is taken with respect to the \ac{irs} configuration
and (from \cite{6404740})
\begin{equation}
V = n' {\rm Var}[I(x;y)] +1 - {\mathbb E}^2\left[\frac{1}{1+h(\phi_k)^2\Lambda_{\rm B}}\right]. 
\end{equation}

% \begin{equation}
% \frac{1}{F} \{ \sum_{k=1}^F
% \end{equation}
%Lastly,
%\begin{equation}\label{eq:Ixz}
%\mathbb{E}[I(x;z)] = I(x;z) = \frac{1}{2}\log_2\left(1 + %\Lambda_{\rm T} \right).
%\end{equation}

Then, we can rewrite \eqref{eq:lai1} and \eqref{eq:lai2} for the H-CRPLA method as, respectively, %transmit a message of size $b_{\rm M}$ bits with decoding error probability $p_{\rm FA}^{\rm (cod-PLA)}$  if}we can transmit a message of size $b_{\rm M}$ bits with decoding error probability $p_{\rm FA}^{\rm (cod-PLA)}$  if}
\begin{subequations}\label{twoeq}
\begin{align}
n'F&\bar{R}  = b_{\rm M} + b_\mathrm{key}, \label{eq:bkm}\\
n'F&[\bar{R} - I(x;z)]  > b_\mathrm{key}. \label{eq:bk}
\end{align}
\end{subequations}
Following the same reasoning as in Section \ref{sec:CDPLA}, we define
\begin{align}\label{eq:bk2}
b_{\rm{key}, 1} &= n'F \bar{R} - b_{\rm M} ,\\
    b_{\rm{key},2} &= n'F\,(\bar{R} - I(x;z) ).    
\end{align}
Then, the number of secret bits of the key is
\begin{equation}
    b_{\rm key} = \max\{0, \min\{b_{\rm{key}, 1} , b_{\rm{key}, 2} \}\}.
\end{equation}

\subsection{Security of Hybrid CR-PLA}

The security of the hybrid \ac{cr}-\ac{pla} mechanism is then equivalent to that provided by a secret key of length
\begin{equation}\label{eq:secbits}
b_{\rm hyb} = b_{\rm ch} + b_{\rm key} ,
\end{equation}
where $b_{\rm ch}$ and $b_{\rm key}$ are given in \eqref{eq:bch} and \eqref{eq:bk2}, respectively.

To design the hybrid \ac{cr}-\ac{pla} mechanism, two trade-offs must be considered: the selection of $\alpha$ and the choice of $h_\mathrm{MIN}$. Indeed, as the value of $\alpha$ increases, the CSI estimate at Bob is more accurate, and $b_{\rm ch}$ increases. However, increasing the number of symbols in a frame dedicated to pilots reduces $b_{\rm key}$. On the other hand, the likelihood of Trudy successfully guessing the key increases when the \ac{snr} on the legitimate channel is low, that is when Bob chooses a configuration $\phi$ leading to a channel gain below the communication-optimal one, i.e., $h(\phi_k)<h_\mathrm{MAX}$.
Therefore, a larger size of the $[h_{\rm MIN}, h_{\rm MAX}]$ interval introduces greater variability in \ac{csi}, thereby increasing $V_{\rm c}$ and consequently $b_{\rm ch}$. However, this comes at the cost of reducing $\mathbb{E}[I_k(x,y)]$, which reduces $b_{\rm key}$.

%introduces more randomness on the resulting \ac{csi} making it more difficult for Trudy to build an attack passing the authentication check. 

%The trade-off between CH-\ac{cr}\ac{pla} and CD-\ac{pla} has to be attained, as increasing the size of the interval $[h_{\rm MIN}, h_{\rm MAX}]$ introduces greater variability in \ac{csi}, thereby increasing $V_{\rm c}$ and consequently $b_{\rm ch}$. However, this comes at the cost of reducing $\mathbb{E}[I(x,y)]$ in \eqref{eq:Ixy}, which reduces $b_{\rm key}$. 

About the \ac{fa} probability, we can simply set $p_{\rm FA}^{CH-CRPLA} =p_{\rm FA}^{CD-PLA} = \frac{p_{\rm FA}}{2}$, where $p_{\rm FA}$ is the overall \ac{fa} probability of H-CRPLA.

\section{Numerical Results}\label{sec:numres}

In this Section, we evaluate the number of secret bits of the (equivalent) key $b_{\rm tot}$, considering $h_{\rm MAX} = 1$, for three authentication mechanisms, thus  
\begin{itemize}
    \item for the \textit{hybrid \ac{cr}-\ac{pla}} (H-CRPLA) mechanism we use $b_{\rm tot} = b_{\rm hyb}$ of \eqref{eq:secbits}, and the parameters $h_\mathrm{MIN}$ and $\alpha$ are chosen to maximize $b_{\rm hyb}$;
    \item the \textit{channel-based \ac{cr}-\ac{pla}} (CH-CRPLA) mechanism, where $b_{\rm tot} = b_{\rm ch}$ with $\alpha = 1$ and $h_\mathrm{MIN}=0$;
    \item the \textit{coding-based \ac{pla}} (CD-PLA) mechanism, where $b_{\rm tot} = b_{\rm key}^{\rm CD}$ is obtained considering the communication-optimal \ac{pcc} configuration and $\alpha = 0$.
\end{itemize}

\begin{figure}
    \centering    \includegraphics[width=\columnwidth]{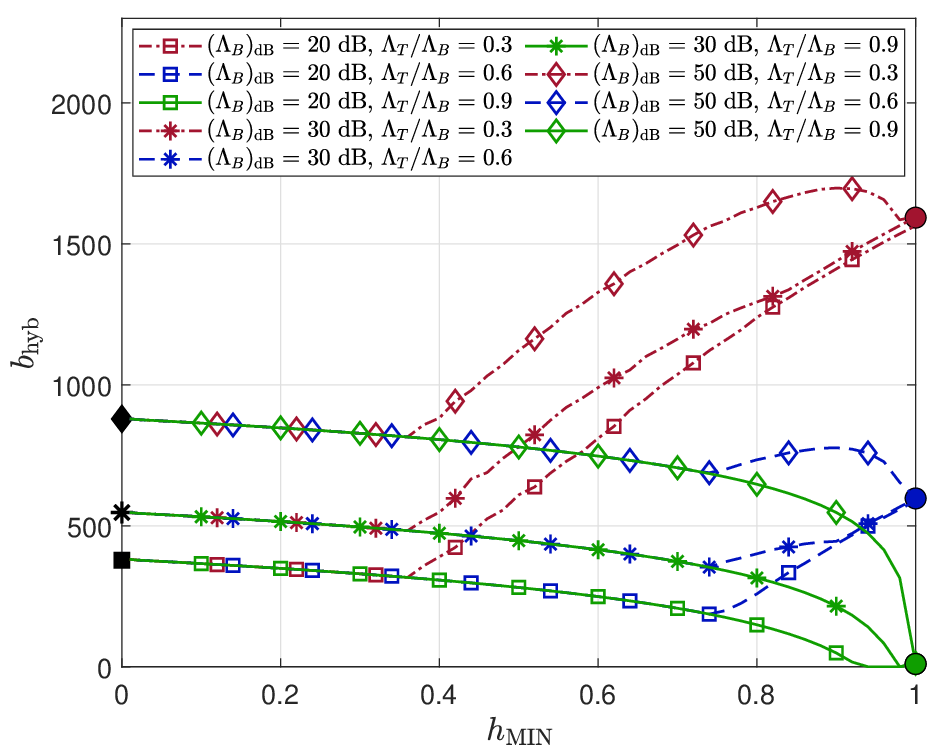}
    \caption{Secret bits $b_{\rm hyb}$ vs $h_\mathrm{MIN}$ for different values of $\Lambda_\mathrm{B}$ and $\Lambda_T/\Lambda_B$, with $F =  100$,  $p_\mathrm{FA} =  10^{-7}$, $n = 10$, and $b_{\rm M} = 600$. The black markers represent the number of secret bits obtained with the CH-CRPLA mechanism for $(\Lambda_{\rm B})_{\rm dB} = $ 20 dB (square), 30 dB (asterisk), and 50 dB (diamond). The colored dots represent the number of secret bits obtained with the CD-PLA mechanism for $\Lambda_T/\Lambda_B = $ 0.3 (red), 0.6 (blue), and 0.9 (green).}
    \label{fig:btotVShmin}
\end{figure}

Fig. \ref{fig:btotVShmin} shows the number of secret bits $b_{\rm hyb}$ obtained with the H-CRPLA method versus $h_\mathrm{MIN}$ for different values of $\Lambda_\mathrm{B}$ and $\Lambda_T/\Lambda_B$, with $F =  100$,  $p_\mathrm{FA} =  10^{-7}$, $n = 10$, and $b_{\rm M} = 600$.
For each value of $h_\mathrm{MIN}$, $\alpha$ is chosen to maximize $b_{\rm hyb}$. As $h_\mathrm{MIN}$ goes to 0, $b_{\rm hyb}$ tends to the number of secret bits obtained with the CH-CRPLA mechanism, represented by the black markers on the left, with different markers for different values of $(\Lambda_{\rm B})_{\rm dB} $. 
As $h_\mathrm{MIN}$ goes to 1, $b_{\rm hyb}$ approaches the number of secret bits obtained with the CD-PLA mechanism, represented by the colored dots on the right of the plot (with different colors for each value of $\Lambda_T/\Lambda_B$).
We note that, as $\Lambda_T/\Lambda_B$ decreases, the CH-CRPLA method permits a higher value of $b_{\rm hyb}$ compared to the CD-PLA method.
Finally, from Fig. \ref{fig:btotVShmin} we note that the proposed H-CRPLA mechanism outperforms both the CD-PLA and the CH-CRPLA when $h_\mathrm{MIN}>0.8$, $(\Lambda_\mathrm{B})_\mathrm{dB} = 50\,$dB, and $\Lambda_T/\Lambda_B = 0.3$.

\begin{figure}
    \centering    \includegraphics[width=\columnwidth]{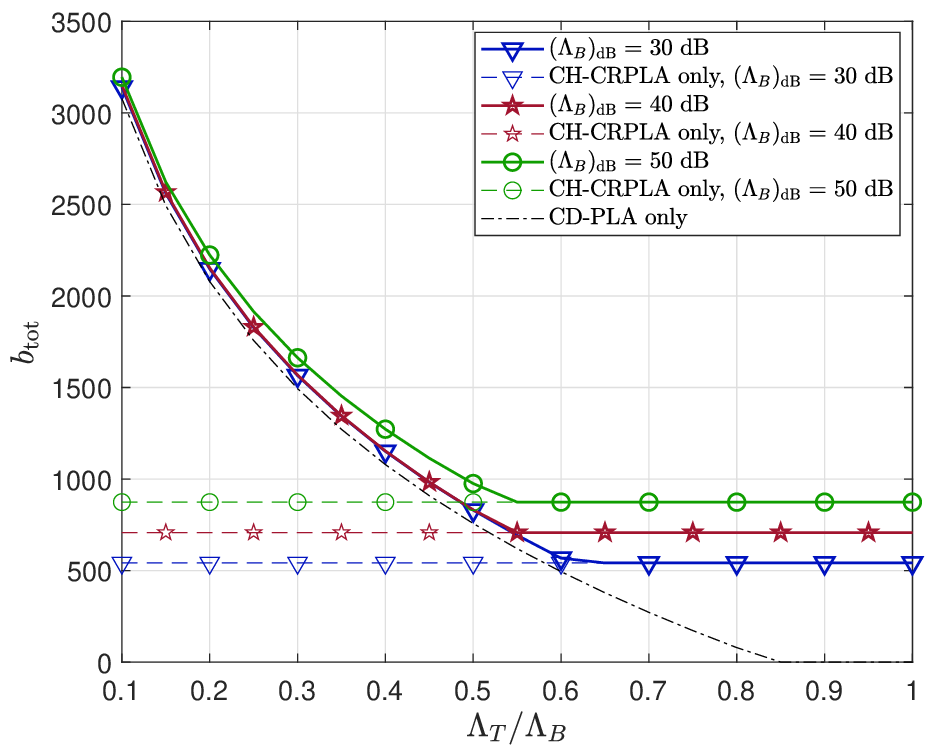}
    \caption{Secret bits $b_{\rm tot}$ vs $\Lambda_T/\Lambda_B$ for different values of $\Lambda_\mathrm{B}$, with $F =  100$,  $p_\mathrm{FA} =  10^{-7}$, $n = 10$, and $b_{\rm M} = 600$.}
    \label{fig:btotVSsnrRatio}
\end{figure}

Fig. \ref{fig:btotVSsnrRatio} shows the number of secret bits $b_{\rm tot}$ achieved with the proposed H-CRPLA mechanism versus $\Lambda_T/\Lambda_B$ for different values of $\Lambda_\mathrm{B}$. 
For $\Lambda_T/\Lambda_B >0.55$, the total number of secret bits for the H-CRPLA mechanism matches that of the CH-CRPLA mechanism. This indicates that adding coding to the CH-CRPLA mechanism does not increase the number of secret bits when $\Lambda_T/\Lambda_B$ is sufficiently high. Moreover, as $\Lambda_\mathrm{B}$ decreases, the number of secret bits for $\Lambda_T/\Lambda_B < 0.55$ tends to that obtained with the CD-PLA method. We observe that the greatest gain in secret bits achieved by the use of the hybrid method occurs when $\Lambda_\mathrm{B}$ is high and $\Lambda_T/\Lambda_B < 0.55$.

\section{Conclusions}\label{sec:concl}
This letter demonstrates that combining coding and \ac{cr} approaches for \ac{pla} strengthens the authentication mechanism. The numerical results validate the effectiveness of this mechanism, showing that the number of secret bits increases with higher \ac{snr} on the legitimate channel and a greater number of frames. These findings suggest that \ac{cr}-\ac{pla} mechanisms, particularly when combined with coding, hold great potential for securing communication in wireless networks.

\balance
\bibliographystyle{IEEEtran}
\bibliography{bibRIS.bib}

\end{document}